\documentclass[]{pasj01}
\draft
\Received{2020/06/dd}
\Accepted{}
 
\usepackage[dvipdfmx]{}
\usepackage{lscape}
\usepackage{threeparttable}
\begin{document} 

\title{ 
FUGIN hot core survey. I. Survey method and initial results for l = 10$^\circ$--20$^\circ$}

\author{Kazuki \textsc{Sato}\altaffilmark{1,2}%
\thanks{Present Address: Komatsu Ltd., 400, Yokokurashinden, Oyama-shi, Tochigi, 323-8558, Japan}}
\altaffiltext{1}{Department of Astronomy, Graduate School of Science, The University of Tokyo, 7-3-1 Hongo, Bunkyo-ku, Tokyo 133-0033, Japan}
\altaffiltext{2}{National Astronomical Observatory of Japan, National Institutes of Natural Sciences, 2-21-1 Osawa, Mitaka, Tokyo 181-8588, Japan}
\altaffiltext{3}{Nobeyama Radio Observatory, National Astronomical Observatory of Japan, National Institutes of Natural Sciences, Minamimaki, Minamisaku, Nagano 384-1305, Japan}
\altaffiltext{4}{Department of Physics, Graduate School of Pure and Applied Sciences, University of Tsukuba, 1-1-1 Tennodai, Tsukuba, Ibaraki 305-8577, Japan}
\altaffiltext{5}{Tomonaga Center for the History of the Universe, University of Tsukuba, 1-1-1 Tennodai, Tsukuba, Ibaraki 305-8571, Japan}
\altaffiltext{6}{Department of Physics, School of Science and Technology, Kwansei Gakuin University, 2-1 Gakuen, Sanda, Hyogo 669-1337, Japan}
\altaffiltext{7}{National Astronomical Observatory of Japan, National Institutes of Natural Sciences, Alonso de C\'{o}rdova 3788, Office 61B, Vitacura, Santiago, Chile}

\email{kazuki.sato.ap@gmail.com}

\author{Tetsuo \textsc{Hasegawa}\altaffilmark{2}}
\author{Tomofumi \textsc{Umemoto}\altaffilmark{3}}
\author{Hiro \textsc{Saito}\altaffilmark{4}}
\author{Nario \textsc{Kuno}\altaffilmark{4,5}}
\author{Masumichi \textsc{Seta}\altaffilmark{6}}
\author{Seiichi \textsc{Sakamoto}\altaffilmark{1,7}}


\maketitle

\begin{abstract}
We have developed a method to make a spectral-line-based survey of hot cores, which represent an important stage of high-mass star formation, and applied the method to the data of the FUGIN  (FOREST Unbiased Galactic plane Imaging survey with the Nobeyama 45-m telescope) survey. 
First, we select hot core candidates by searching the FUGIN data for the weak hot core tracer lines (HNCO and CH$_3$CN) by stacking, and then we conduct follow-up pointed observations on these candidates in C$^{34}$S, SO, OCS, HC$_3$N, HNCO, CH$_3$CN, and CH$_3$OH $J=2-1$ and $J=8-7$ lines to confirm and characterize them. 
We applied this method to the $l$ = 10$^\circ$--20$^\circ$ portion of the FUGIN data and identified 22 ``Hot Cores'' (compact sources with more than two significant detection of the hot core tracer lines, i.e., SO, OCS, HC$_3$N, HNCO, CH$_3$CN, or CH$_3$OH $J=8-7$ lines) and 14 ``Dense Clumps'' (sources with more than two significant detection of C$^{34}$S, CH$_3$OH $J=2-1$, or the hot core tracer lines). 
The identified Hot Cores are found associated with signposts of high-mass star formation such as ATLASGAL clumps, WISE H\,\emissiontype{II} regions, and Class II methanol masers. 
For those associated with ATLASGAL clumps, their bolometric luminosity to clump mass ratios are consistent with the star formation stages centered at the hot core phase. 
The catalog of FUGIN Hot Cores provides a useful starting point for further statistical studies and detailed observations of high-mass star forming regions.
\end{abstract}

\setcounter{page}{3}
\newpage

\section{Introduction}

High-mass stars play important roles in determination of their environments in galactic scale. They irradiate their surrounding gas and dust by strong UV light to heat and ionize them and drive stellar winds throughout their lifetime, and they finally explode as supernovae to make significant impacts even after the end of their lives. However, forming process of high-mass stars is still under debate. It is important to study the early stages of their formation process.

Often observed in high-mass star formation regions are ``hot cores'', which are hot ($>$ 100 K) and compact ($<$ 0.1 pc) cores of dense ($>10^7$ cm$^{-3}$) molecular gas with large extinction ($A_{\rm v} >100$ mag) and characteristic chemistry (e.g., \cite{nom04}).
In cool molecular clouds, gas-phase molecules are adsorbed on the surface of dust grains to form mantle. Chemical reaction occurs in both mantle around the dust grains and gas phase. Molecules forming in dust mantle include complex organic molecules (COMs) (e.g., \cite{yam17}). 

When high-mass stars are formed, surrounding dust grains are irradiated and become hot. Molecules in dust mantle evaporate to gas phase. Hot cores are observed with the characteristic emission lines of molecules that evaporate from the grains.

One of the key questions in researches of high-mass star formation is the conditions to form high-mass stars.
Statistical studies of high-mass star forming regions are essential to address this issue. Surveys of high-mass star formation regions are mainly conducted in dust continuum emission. 
For instance, toward the clumps identified by the APEX Telescope Large Area Survey of the Galaxy (ATLASGAL), which is an 870 $\mu$m survey of the Galactic plane, spectral-line follow-ups have been made (\cite{urq18}). 
Similarly, search for dense gas in CS $J=2-1$ emission was conducted toward IRAS point sources as an H\,\emissiontype{II} region survey (\cite{bro95}). 
H\,\emissiontype{II} region catalogs of the Galactic plane were made using mid-infrared emission detected by the WISE satellite (\cite{and14}) and by selecting compact sources in 70 $\mu$m from Hi-GAL (Herschel infrared Galactic Plane Survey) source catalog (\cite{mol16}). 

These continuum-based surveys have, however, potential weak points. First, the millimeter/submillimeter dust continuum emission is roughly proportional to the product of dust column density and dust temperature. The continuum-based surveys thus naturally pick more massive cores relative to the less massive hot cores because of large range of dust column density than dust temperature. Secondly, source confusion can limit the separation of cores in crowded areas such as cluster forming regions because continuum observations cannot separate sources by radial velocities.

To overcome the above weakness, it is important to conduct an unbiased survey in emission lines of hot core tracer molecules.
Here we utilize the archived data of FOREST Unbiased Galactic plane Imaging survey with the Nobeyama 45-m telescope (FUGIN), which is a survey of the Galactic plane in $l = 10^\circ$--$50^\circ$ and $198^\circ$--$236^\circ$, $|b| \leq 1^\circ$ using $^{12}$CO, $^{13}$CO and C$^{18}$O $J=1-0$ lines with additional lines in the frequency band. The velocity resolution of FUGIN is 1.3 km s$^{-1}$. The beam size is $14''$ and the angular resolution is $21''$ for $^{13}$CO and C$^{18}$O (\cite{ume17}).
Since FUGIN is the Galactic plane CO survey made at the highest spatial resolution so far, it is most suited to find compact sources such as hot cores. Although FUGIN was targeted on the CO $J=1-0$ lines, the observed frequency band included some other weaker molecular lines such as HNCO $J_{K_a, K_c}=5_{0, 5}-4_{0, 4}$ (109.906 GHz) and CH$_3$CN $J_K=6_K-5_K$\ ($K = 0, 1, 2, 3$; 110.364--110.384 GHz), which are known as good tracers of hot cores. We used these lines to survey hot cores without bias to the continuum emission. This research is the first approach of hot core survey using hot core tracer lines. We can discuss statistical characteristics of hot cores without bias to bright continuum emission. 

In this paper, we establish a method to select candidate hot cores from the survey data, followed by confirmation observations (section 2), and introduce initial results for the $l = 10^\circ$--$20^\circ$ area (section 3). In section 4, we characterize the nature of the detected sources. 

\section{Method}\label{s2}

Our hot core survey based on the FUGIN data takes two major steps. 
First, we search for candidates of hot cores by looking at the C$^{18}$O, HNCO, and CH$_3$CN lines observed in the FUGIN survey. Although FUGIN observations were made as a survey of $^{12}$CO, $^{13}$CO and C$^{18}$O $J=1-0$ lines, the observed frequency bands included some other molecular lines.
Then, we make pointed observations towards the candidate sources at higher sensitivity to confirm and characterize the sources. 
In the following, we describe these steps in detail.

\subsection{Candidate selection}

\begin{figure}
 \begin{center}
  \includegraphics[width=8cm]{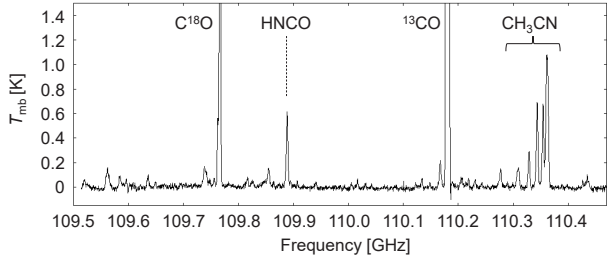} 
 \end{center}
\caption{Spectrum near the $^{13}$CO and C$^{18}$O $J=1-0$ lines of W51 e1/e8 taken during the FUGIN survey for calibration purposes. }\label{fig:e1e8}
\end{figure}

First, we analyze the FUGIN survey data (\cite{ume17}) to search for hot core candidates. 
Figure \ref{fig:e1e8} shows the spectrum of the hot core source W51 e1/e8, which was observed repeatedly during the FUGIN survey for calibration purposes. 
In this high-sensitivity spectrum, we clearly see the emission lines of HNCO $J_{K_a, K_c}=5_{0,5}-4_{0,4}$ and CH$_3$CN $J_K=6_K-5_K$ ($K = 0, 1, 2, 3, 4$) in addition to the $^{13}$CO and C$^{18}$O $J=1-0$ lines. 
These HNCO and CH$_3$CN lines are known as tracers of hot cores (e.g., \cite{bis07}).

The sensitivity of the actual FUGIN survey has been set for the $^{12}$CO, $^{13}$CO and C$^{18}$O lines, and the resultant rms noise is d$T_{\rm mb} \approx$ 0.2--0.4 K for the $^{13}$CO and C$^{18}$O lines.
This makes it difficult to detect hot core tracer lines from the survey data if we depend on only one line of hot core tracer molecule. 
To improve the signal-to-noise ratio, we stacked the line of HNCO $J_{K_a, K_c}=5_{0,5}-4_{0,4}$ and four lines of CH$_3$CN $J_K=6_K-5_K$ ($K = 0, 1, 2, 3$). 

In practice, this is still insufficient to efficiently exclude false detections from candidate sources for confirmation observations.
Through our preliminary analysis in W51 e1/e8, we found that real sources were always associated with relatively bright C$^{18}$O emission ($T_{\rm mb}$(C$^{18}$O) $ \geq 1.5$ K), which indicates regions of high column density of molecular gas. 
Since we can safely assume that hot cores exist in regions of high molecular column density, we require the hot core candidates to be associated with bright C$^{18}$O emission. 
We thus selected our candidates in the following three steps: 
\begin{enumerate}
\item In the C$^{18}$O datacube, we identify positions with $T_{\rm mb}$(C$^{18}$O) $\geq$ 1.5 K and record the corresponding peak velocities. By assuming the conditions typical for dense gas, this roughly corresponds to $N{\rm(H_2)} \geq 8 \times 10^{23}$ cm$^{-2}$.
\item We calculate the integrated intensity of the stacked HNCO and CH$_3$CN lines over $\pm$5 km s$^{-1}$ centered at the peak velocity of C$^{18}$O. 
If the integrated intensity exceeds 5$\sigma$, we regard it as a positive detection. 
\item When we find more than two adjacent positive detections in the datacube, we regard them as a single source at the position where we get the strongest integrated intensity of the stacked lines. 
\end{enumerate}

\begin{figure}
 \begin{center}
  \includegraphics[width=8cm]{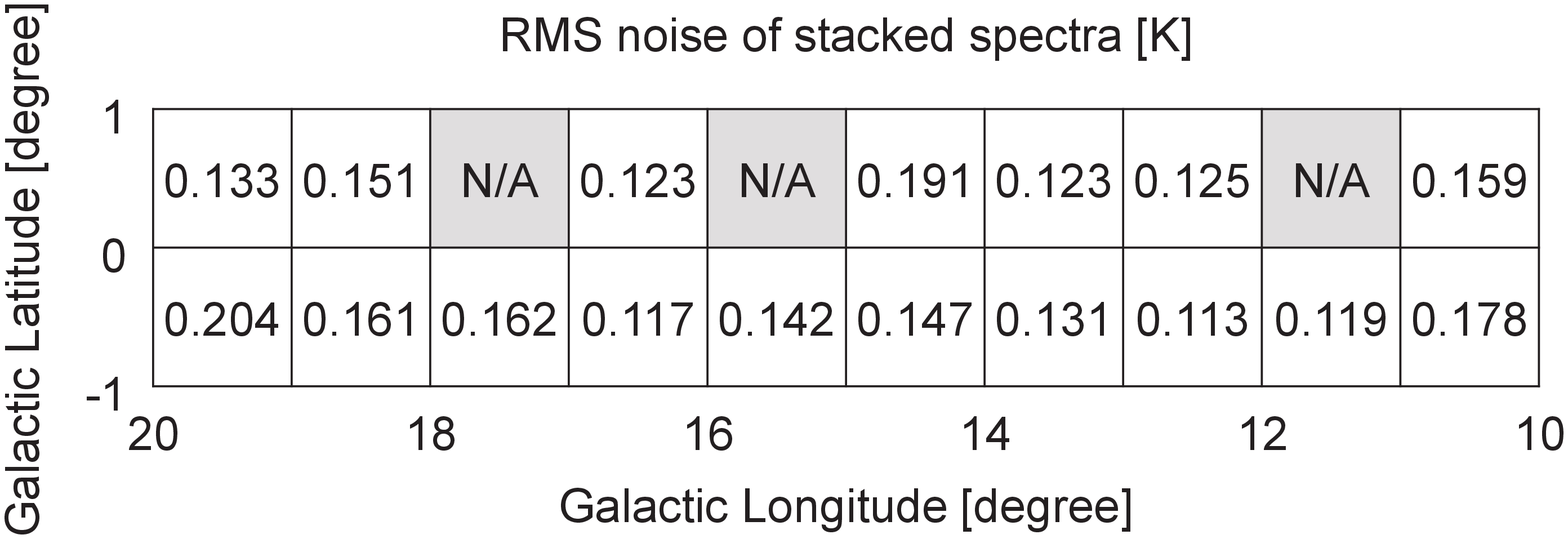} 
 \end{center}
\caption{The rms noise levels of the stacked specra at 110 GHz in each squared degrees at 1.3 km s$^{-1}$ velocity resolution. 
In three regions (12$^\circ \leq l \leq $13$^\circ$, 0$^\circ \leq b \leq $1$^\circ$; 15$^\circ \leq l \leq $16$^\circ$, 0$^\circ \leq b \leq $1$^\circ$; and 17$^\circ \leq l \leq $18$^\circ$, 0$^\circ \leq b \leq $1$^\circ$), there are strong scanning effect and we did not extract hot core candidates from these regions.}
\label{fig:rms_map}
\end{figure}

We applied this method to the $10^\circ \leq l \leq 20^\circ$, $-1^\circ \leq b \leq 1^\circ$ part of the FUGIN data except the three square degrees regions ($12^\circ \leq l \leq 13^\circ$, $0^\circ \leq b \leq 1^\circ$; $15^\circ \leq l \leq 16^\circ$, $0^\circ \leq b \leq 1^\circ$; and $17^\circ \leq l \leq 18^\circ$, $0^\circ \leq b \leq 1^\circ$), which have strong scanning noise and are not suitable for the candidate selection. An rms level of each squared degree is 0.1--0.2 K in the $T_{\rm mb}$ scale as shown in figure \ref{fig:rms_map}. We adopted this rms noise levels to calculate the candidate selection threshold in each squared degrees.
Consequently, we analyzed the 17 square degrees area of the FUGIN survey following the procedure described above and found 64 candidates for confirmation.

\subsection{Confirmation observation}

We focus on characteristic chemical abundance of hot core to identify the sources.
In addition, we investigate compactness of the molecular distribution to exclude shock-originated molecular lines.

We conducted the observations to hot core candidates using the Nobeyama 45-m radio telescope in 2018 March and May. The 4-beam receiver FOREST (\cite{min16}) and the autocorrelation spectrometer SAM45 (\cite{kun11}) were used. 
The RF frequency ranges were 94.8--96.8 GHz and 108.8--110.8 GHz in the lower and upper sideband, respectively. 
The spectrometer was configured to cover 4 GHz in total at a 448.28 kHz channel separation.
The corresponding velocity resolution varies from 1.52 km s$^{-1}$ at 96 GHz to 1.33 km s$^{-1}$ at 110 GHz. 
The beam size is $17''$ at 96 GHz and $14''$ at 110 GHz. 
The system noise temperatures were 150--350 K at 96 GHz and 200--400 K at 110 GHz during the observations. 
SiO masers were observed every 1 hour to calibrate the pointing offset of the telescope, and the pointing error was typically less than $3''$.

The major spectral lines in the observed frequency ranges are listed in table \ref{tab:lines}. 
In addition to the C$^{18}$O $J=1-0$, HNCO $J_{K_a, K_c}=5_{0, 5}-4_{0, 4}$ and CH$_3$CN $J=6-5$ lines that we used to find the candidates, we observed C$^{34}$S $J=2-1$, SO $J_K=2_3-1_2$, OCS $J=9-8$, HC$_3$N $J=12-11$, and CH$_3$OH lines, which are often quoted as tracers of hot cores. 
For CH$_3$OH, in addition to the $J_K=2_{0}-1_{0} $ A$^+$, $J_K=2_{-1}-1_{-1} $ E  and $J_K=2_0-1_0 $ E lines, we observed the $J_K=8_0-7_1$ A$^+$ line, which is known to exhibit maser action in some high-mass star forming regions (e.g., \cite{che11}). 

\begin{table*}
  \tbl{List of important molecular lines included in the confirmation observation. \\}{
  \begin{tabular}{cccc}
      \hline
      Molecule & Transition & Frequency [GHz] & $E_{\rm u}^*/k$ [K]    \\ 
      \hline
      C$^{18}$O & $J=1-0$ & 109.782173 & \phantom{00}5.27 \\
      C$^{34}$S  & $J=2-1$& \phantom{0}96.412950 & \phantom{00}6.94 \\
      SO            & $J_K=2_3-1_2$ & 109.252220 & \phantom{0}21.05 \\
      OCS         & $J=9-8$ & 109.463063 & \phantom{0}26.26 \\
      HC$_3$N   & $J=12-11$ & 109.173634 & \phantom{0}34.06 \\
      HNCO       & $J_{K_a, K_c}=5_{0,5}-4_{0,4}$ & 109.905749 & \phantom{0}15.82 \\
      CH$_3$CN & $J_K=6_0-5_0$ & 110.383500 & \phantom{0}18.54 \\
      CH$_3$CN & $J_K=6_1-5_1$ & 110.381372 & \phantom{0}25.68 \\
      CH$_3$CN & $J_K=6_2-5_2$ & 110.374989 & \phantom{0}47.13 \\
      CH$_3$CN & $J_K=6_3-5_3$ & 110.364354 & \phantom{0}82.85 \\
      CH$_3$CN & $J_K=6_4-5_4$ & 110.349470 &132.84 \\
      CH$_3$OH & $J_K=2_0-1_0$ A$^+$ & \phantom{0}96.741371 & \phantom{00}6.96 \\
      CH$_3$OH & $J_K=2_{-1}-1_{-1}$ E & \phantom{0}96.739358 & \phantom{0}12.54 \\
      CH$_3$OH & $J_K=2_0-1_0$ E & \phantom{0}96.744545 & \phantom{0}20.09 \\
      CH$_3$OH & $J_K=8_0-7_1$ A$^+$ & \phantom{0}95.169391 & \phantom{0}83.54 \\
      \hline
$*$ Upper state energy.\\
    \end{tabular}}\label{tab:lines}
\end{table*}

\begin{figure}
 \begin{center}
  \includegraphics[width=5cm]{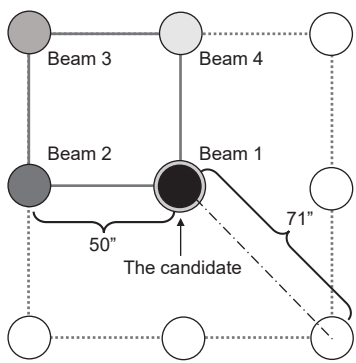} 
 \end{center}
\caption{The mapping pattern of the confirmation observation with the FOREST four beams. We observed the candidate at the center with Beam 1, 2, 3 and 4 in turn, so that we can get spectra of the positions $50''$ and $71''$ away from the center.}\label{fig:four_beam}
\end{figure}

We observed the 64 candidates in position switching mode using the FOREST four-beam receiver with the pattern shown in figure \ref{fig:four_beam}.
The central position of a candidate was observed by the four beams in turn. 
As a result, we got a $3 \times 3$ map with a $50''$ grid centered at each candidate. 
This allows us to judge if the emission is spatially confined ($\ll 100''$ = 1.45 pc at 3 kpc) or not, and to exclude extended objects like shocked gas from the hot core candidate list. 
The integration time was set to achieve a target sensitivity of  d$T_{\rm mb} \approx$ 0.03--0.05 K, which was $\sim$ 10 times lower than that of the original FUGIN survey.

\section{Results}

\subsection{Identification of Hot Cores and Dense Clumps}

\begin{figure*}
 \begin{center}
\includegraphics[width=16cm]{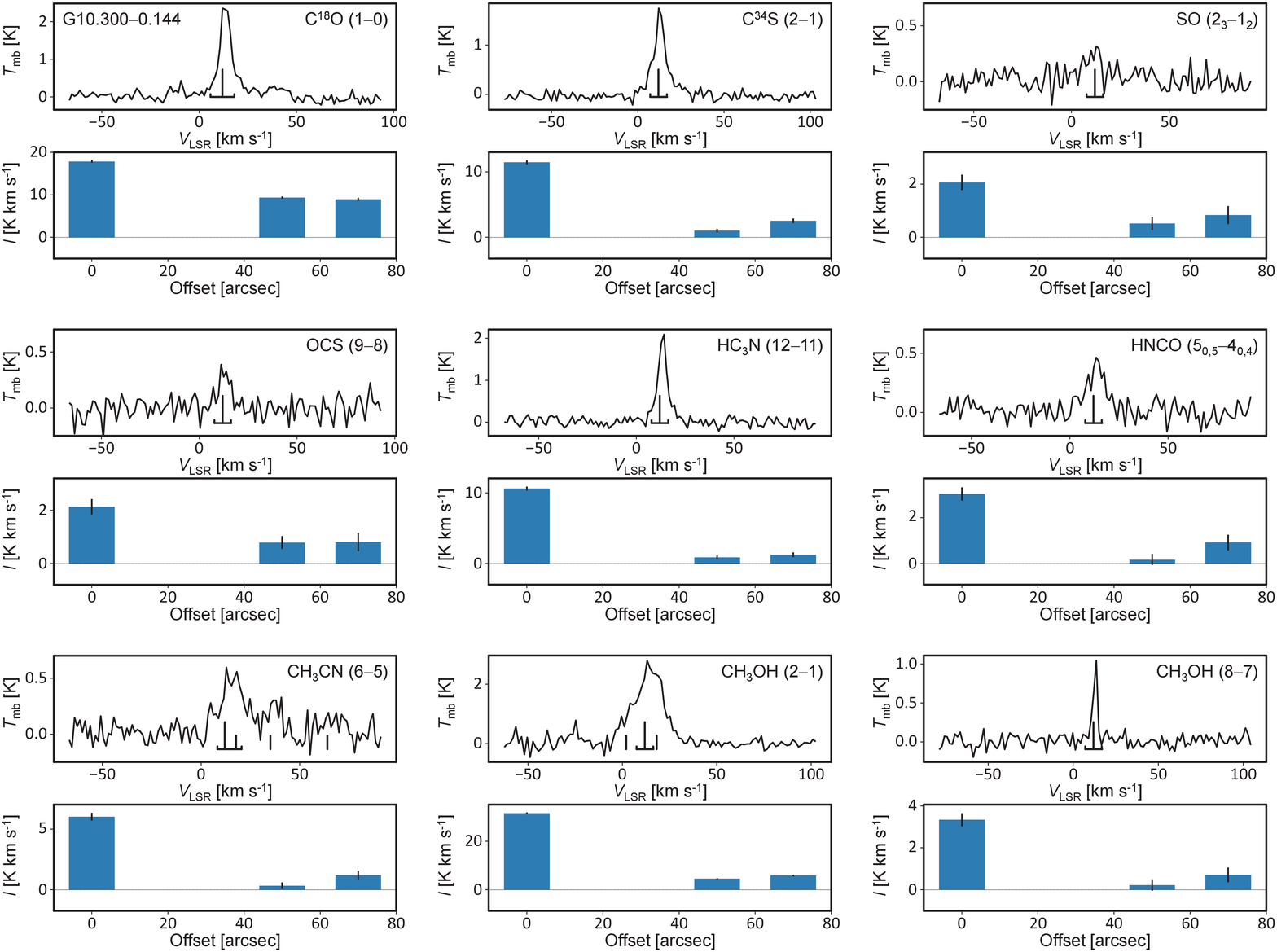} 
 \end{center}
\caption{An example of the results of the confirmation observations 
showing the profiles of nine molecular lines observed from G10.300$-$0.144. 
Upper part of each panel shows the line profile. 
The horizontal axis is the radial velocity, $V_{\rm LSR}$ (km s$^{-1}$), 
and the vertical axis is the main beam antenna temperature, $T_{\rm mb}$ (K). 
The long vertical lines indicate the peak velocity of the C$^{18}$O line, 
and the short vertical lines for the CH$_3$CN and CH$_3$OH lines 
correspond to the frequency of their weaker lines. 
The horizontal lines with ticks on both ends indicate the velocity range of integration.
Lower side of each panel shows the integrated intensity of the line at the center, 
$I = \int T_{\rm mb}\ {\rm d}V$ (K km s$^{-1}$), and the average intensities 
at positions $50''$ and $71''$ away from the center. We can judge 
the compactness of the emission region by comparison of these three intensities.}
\label{fig:G1300}
\end{figure*}

Figure \ref{fig:G1300} shows an example of the results of our follow-up observations. 
It shows the line profiles (upper panels) and integrated intensity distributions (lower panels) of the nine important molecular lines observed from one of identified hot core sources, G10.300$-$0.144. 
For each molecular line, we see its line profile and its spatial extent in a bar graph showing the integrated intensities at the center and the positions $50''$ and $71''$ away from the center. 
The intensities at $50''$ and $71''$ offsets are averages of the intensities at the four positions, i.e., north, south, east and west in the Galactic coordinate at $50''$ offset and northeast, northwest, southeast and southwest at $71''$ offset, respectively (figure \ref{fig:four_beam}). 
The 1$\sigma$ errors are shown by the solid lines at the top of the bars. 
The errors shown for the offset positions represent only the statistical errors and do not include the intrinsic intensity variation among the four positions in the average. 
With exceptions of C$^{18}$O $J=1-0$ and CH$_3$CN lines, the line intensities are integrated over seven velocity channels ($V_{\rm peak} - 4.9$ km s$^{-1}  \leq V_{\rm LSR} \leq  V_{\rm peak} + 4.9$ km s$^{-1}$ at 96 GHz and $V_{\rm peak} - 4.3$ km s$^{-1}  \leq V_{\rm LSR} \leq  V_{\rm peak} + 4.3$ km s$^{-1}$ at 110 GHz) centered at the peak velocity of the C$^{18}$O emission ($V_{\rm peak}$). 
The C$^{18}$O $J=1-0$ line is integrated over 10 channels ($V_{\rm peak} - 6.1$ km s$^{-1}  \leq V_{\rm LSR} \leq  V_{\rm peak} + 6.1$ km s$^{-1}$). 
In the case of CH$_3$CN, the strongest $J_K=6_0-5_0$ line is blended with the neighboring $J_K=6_1-5_1$ line, and wider velocity range is used for integration ($V_{\rm peak} - 3.7$ km s$^{-1}  \leq V_{\rm LSR} \leq  V_{\rm peak} + 8.5$ km s$^{-1}$) to include the $J_K=6_1-5_1$ line. 

We regard a line is detected if the integrated intensity exceeds 3$\sigma$ and its profile has a reasonable shape centered at the source velocity defined by the C$^{18}$O line. 
The spatial compactness of the molecular emission is judged from a comparison of the integrated intensity at the center with those $50''$ and $71''$ away. 
At the distance of 3 kpc, $50''$ corresponds to 0.73 pc, which is significantly larger than the typical size of hot cores, detection of hot core tracer lines is expected only toward the center. 
In the example shown in figure \ref{fig:G1300}, all of the nine molecular lines are detected, and their spatial distribution is compact except for C$^{18}$O. 

The confirmation observation results of the 64 candidates show a range of significance for identification as a hot core. 
While we detect many of the spectral lines in table \ref{tab:lines} from some candidates, other sources only show the C$^{34}$S and CH$_3$OH $J=2-1$ lines besides C$^{18}$O. 
Based on the detected lines and their spatial distributions, we classify the candidates into three categories; ``Hot Cores'', ``Dense Clumps'', and non-detections with the following definitions\footnote{Since the definitions of the terms ``Hot Cores'' and ``Dense Clumps'' used in this paper can be slightly different from the classical definitions, we write them in our definitions with capital letters.}.

\begin{description}
\setlength{\parskip}{0 cm}
\setlength{\itemsep}{0 cm}
\item[Hot Cores]~\\
The sources with detection of at least two of the hot core lines (SO, OCS, HC$_3$N, HNCO, CH$_3$CN and CH$_3$OH $J=8-7$) with clear spatial concentration at the center judged from the comparison of the line intensity at the center with those at positions $50''$ and $71''$ away from the center. 
G10.300$-$0.144 shown in figure \ref{fig:G1300} is a typical example of this category.
\item[Dense Clumps]~\\
The sources that fail to meet the Hot Core criteria, but with detection of at least two of C$^{34}$S, CH$_3$OH $J=2-1$, or the hot core lines listed above.
\item[Non-detections]~\\
The candidates that do not meet the criteria for the above two categories are not considered as positive detection. 
Most of them may represent the inevitable false positive detections in the candidate selection step. 
\end{description}

We identified 22 Hot Cores and 14 Dense Clumps, which represent 34\% (Hot Cores) and 22\% (Dense Clumps) of the observed 64 candidates, respectively. 
Table \ref{tab:hc_cat} shows the identified Hot Cores and Dense Clumps according to the criteria above. 
Method to estimate distances of each candidate is described in section 3.3. 
The intensities with the parentheses in table \label{tab:hc_cat} are those below the 3$\sigma$ detection thresholds. 
Many of the Dense Clumps are detected only in C$^{34}$S and CH$_3$OH $J=2-1$. 
We classify G10.203$-$0.341, G10.209$-$0.336, and G12.816$-$0.185 as Dense Clumps despite the multiple detection of hot core lines, because their spatial distributions of the line intensities are not concentrated. 

We note that the criteria employed in this work to identify Hot Cores do not explicitly include direct indices of high gas temperature (e.g., $\sim$ 100 K or higher). 
In this respect, the definition of Hot Cores in this paper may be broader than the classical definition, and the Hot Cores identified here can include also the warm envelopes with similar chemical characteristics that surround forming high-mass stars. 
An analysis of the excitation temperatures of CH$_3$CN and CH$_3$OH is being planned to further explore this point, but it is outside of the scope of the present paper. 

\begin{table*}
\caption{Identified Hot Cores and Dense Clumps with integrated intensities\\}
\scriptsize
  \begin{tabular}{ccc|ccccccccc}  
\hline
 & & & \multicolumn{9}{c}{Integrated intensity [K km s$^{-1}$]}  \\ 
Name & $V_{\rm{LSR}}$ & Distance\footnotemark[$*$] & C$^{18}$O & C$^{34}$S & SO & OCS & HC$_3$N & HNCO & CH$_3$CN & CH$_3$OH & CH$_3$OH \\
 & [km s$^{-1}$] & [kpc] & $1-0$ & $2-1$ & $2_3-1_2$ & $9-8$ & $12-11$ & $5_{0,5}-4_{0,4}$ & $6-5$\footnotemark[$\dag$] & $2-1$\footnotemark[$\ddag$] & $8-7$\footnotemark[$\ddag$] \\ 
\hline
{\bf Hot Cores} & & & & & & & & & & & \\ 
G10.145$-$0.339 & 10.6 & 3.12 $\pm$0.21 & 13.49 & 2.92 & 1.33 & ($-0.12$) & (0.02) & 1.11 & (0.66) & 4.88 & (0.17) \\
G10.178$-$0.350 & 13.3 & 3.12 $\pm$0.21 & 23.29 & 6.49 & (0.39) & (0.62) & 3.93 & 1.92 & (1.11) & 7.14 & (0.60) \\
G10.187$-$0.344 & 12.0 & 3.12 $\pm$0.21 & 22.65 & 2.09 & (0.67) & 1.31 & 1.85 & 2.91 & (1.17) & 11.22 & (0.22) \\
G10.220$-$0.369 & 12.0 & 3.12 $\pm$0.21 & 26.02 & 1.89 & (0.27) & (0.77) & 1.43 & 2.43 & ($-0.11$) & 8.53 & (0.09) \\
G10.300$-$0.144 & 12.0 & 3.13 $\pm$0.22 & 17.84 & 11.46 & 2.07 & 2.14 & 10.61 & 3.03 & 6.03 & 31.53 & 3.34 \\
G10.465$+$0.034 & 72.0 & 8.55 $^{+\,0.63}_{-\,0.55}$ & 11.21 & 2.27 & 0.84 & (0.49) & 1.83 & 2.99 & 0.87 & 12.40 & (0.49) \\
G10.473$+$0.028 & 66.6 & 8.55 $^{+\,0.63}_{-\,0.55}$ & 24.07 & 27.50 & 11.79 & 13.28 & 22.52 & 14.94 & 32.48 & 66.01 & 14.27 \\
G10.481$+$0.036 & 66.6 & 8.55 $^{+\,0.63}_{-\,0.55}$ & 10.92 & 2.96 & 1.33 & 1.51 & 3.96 & 6.79 & 2.92 & 29.96 & 2.41 \\
G10.623$-$0.380 & $-2.7$\phantom{:} & 4.95 $^{+\,0.51}_{-\,0.43}$ & 57.17 & 33.82 & 7.01 & 3.98 & 25.56 & 5.66 & 11.25 & 39.99 & 7.84 \\
G11.920$-$0.616 & 36.0 & 3.37 $^{+\,0.39}_{-\,0.32}$ & 12.80 & 4.55 & 1.24 & 2.41 & 8.63 & 6.12 & 6.30 & 39.33 & 8.76 \\
G11.939$-$0.616 & 38.6 & 3.37 $^{+\,0.39}_{-\,0.32}$ & 18.32 & 7.39 & 0.65 & (0.74) & 11.59 & 1.62 & 3.77 & 14.76 & 1.72 \\
G12.416$+$0.510 & 18.6 & 1.83 $\pm$0.09 & 12.66 & 3.46 & 1.10 & 1.10 & 4.30 & (0.69) & (0.66) & 9.49 & (0.70) \\
G12.680$-$0.176 & 56.0 & 2.40 $^{+\,0.17}_{-\,0.15}$ & 26.91 & 2.32 & (0.37) & ($-0.04$) & 2.47 & 1.49 & (0.43) & 12.78 & (0.45) \\
G12.810$-$0.204 & 34.6 & 2.92 $^{+\,0.35}_{-\,0.28}$ & 48.73 & 27.96 & 3.95 & 2.27 & 20.78 & 3.42 & 5.70 & 28.83 & 2.88 \\
G12.888$+$0.490 & 33.3 & 2.50 $^{+\,0.28}_{-\,0.23}$ & 27.43 & 11.42 & 3.68 & 2.91 & 10.92 & 2.30 & 5.38 & 12.96 & 6.41 \\
G13.185$-$0.109 & 53.3 & 4.60 $\pm$0.23 & 14.81 & 1.27 & (0.26) & ($-0.18$) & 2.41 & 1.64 & 1.54 & 11.19 & (0.38) \\
G13.199$-$0.134 & 52.0 & 4.60 $\pm$0.23 & 16.17 & 1.02 & (0.10) & (0.42) & 0.81 & 1.55 & ($-0.37$) & 10.03 & ($-0.22$) \\
G13.216$+$0.029 & 52.0 & 4.57 $\pm$0.22 & 14.66 & 1.00 & (0.11) & 0.80 & 1.29 & 1.82 & 0.93 & 13.71 & 0.95 \\
G14.113$-$0.569 & 21.3 & 1.86 $\pm$0.10 & 15.24 & 2.13 & 1.60 & (0.18) & 1.33 & (0.67) & 1.56 & 5.02 & (0.60) \\
G14.330$-$0.644 & 21.3 & 1.12 $^{+\,0.14}_{-\,0.11}$ & 18.20 & 12.07 & 8.60 & 4.40 & 22.03 & 3.41 & 9.52 & 50.81 & 57.76 \\
G14.636$-$0.569 & 18.6 & 1.83 $^{+\,0.08}_{-\,0.07}$ & 9.64 & 2.65 & 0.48 & (0.30) & 2.77 & 0.91 & 0.91 & 15.57 & 1.43 \\
G19.367$-$0.041 & 26.6 & 3.11 $\pm$0.26 & 14.47 & 2.62 & 0.48 & 1.00 & 0.56 & (0.24) & ($-0.13$) & 10.05 & (0.57) \\ 
\hline
{\bf Dense Clumps} & & & & & & & & & & & \\ 
G10.131$-$0.408 & 10.6 & 3.12 $\pm$0.21 & 12.69 & 0.96 & (0.00) & ($-0.48$) & (0.12) & (0.56) & ($-0.28$) & 4.60 & ($-0.08$) \\
G10.195$-$0.397 & 10.6 & 3.12 $\pm$0.21 & 20.23 & 1.74 & ($-0.23$) & ($-0.03$) & 1.12 & (0.74) & (0.03) & 5.83 & (0.04) \\
G10.203$-$0.341 & 12.0 & 3.12 $\pm$0.21 & 24.38 & 1.79 & ($-0.06$) & (0.90) & (1.07) & 0.98 & 0.58 & 7.02 & (0.43) \\
G10.203$-$0.308 & 10.6 & 3.12 $\pm$0.21 & 18.60 & (0.93) & (0.26) & (0.18) & (0.43) & 1.54 & (1.10) & 2.08 & (0.54) \\
G10.209$-$0.336 & 12.0 & 3.12 $\pm$0.21 & 28.21 & 1.73 & (0.33) & (0.23) & 1.67 & 2.04 & (0.32) & 7.84 & (0.61) \\
G10.631$-$0.372 & $-1.4$\phantom{:} & 4.95 $^{+\,0.51}_{-\,0.43}$ & 17.46 & 3.25 & ($-0.21$) & (0.43) & 2.41 & (0.79) & (0.15) & 5.73 & (0.75) \\
G11.059$-$0.372 & $-1.4$\phantom{:} & 4.91 $\pm$0.25 & 12.70 & (0.32) & (0.26) & (0.37) & ($-0.06$) & 0.83 & (0.14) & 2.07 & ($-0.25$) \\
G11.109$-$0.388 & \phantom{0}0.0 & 4.91 $\pm$0.25 & 10.83 & 0.55 & ($-0.12$) & (0.14) & (0.10) & (0.31) & (0.15) & 2.82 & ($-0.02$) \\
G12.685$-$0.171 & 54.6 & 2.40 $^{+\,0.17}_{-\,0.15}$ & 18.89 & 1.19 & ($-0.09$) & (0.00) & 0.74 & 0.86 & (0.01) & 6.18 & (0.22)\\
G12.816$-$0.185 & 34.6 & 2.92 $^{+\,0.35}_{-\,0.28}$ & 19.23 & 5.33 & (0.35) & 0.72 & 4.52 & 2.37 & 2.31 & 24.47 & 1.09 \\
G12.855$-$0.201 & 36.0 & 2.92 $^{+\,0.35}_{-\,0.28}$ & 17.08 & 1.16 & ($-0.14$) & ($-0.21$) & 0.78 & (0.41) & (0.70) & 2.37 & ($-0.13$) \\
G12.911$+$0.487 & 32.0 & 2.50 $^{+\,0.28}_{-\,0.23}$ & 13.81 & 1.64 & ($-0.54$) & (0.00) & (0.81) & (0.79) & ($-1.07$) & 3.35 & (0.64) \\
G13.202$+$0.057 & 52.0 & 4.57 $\pm$0.22 & 12.17 & 1.13 & (0.16) & ($-0.06$) & 0.88 & (0.68) & (0.82) & 2.35 & (0.14) \\
G14.202$-$0.199 & 38.6 & 3.09 $\pm$0.39 & 13.02 & 1.55 & ($-0.10$) & (0.04) & (0.14) & (0.34) & (0.93) & 8.88 & (0.11) \\
\hline
  \end{tabular}
  \label{tab:hc_cat}
\footnotemark[$*$] See section 3.3 for the methods of distance determination. \\
\footnotemark[$\dag$] This column shows the integrated intensities of the blend of the CH$_3$CN $J_K=6_0-5_0$ and $J_K=6_1-5_1$ lines. \\
\footnotemark[$\ddag$] The column of CH$_3$OH $J=2-1$ shows the integrated intensities of the blend of the $J_K=2_{0}-1_{0}$ A$^+$ and $J_K=2_{-1}-1_{-1}$ E lines, and the column of CH$_3$OH $J=8-7$ shows the integrated intensities of the $J_K=8_0-7_1$ A$^+$ line. \\
\end{table*}

\subsection{Clustering of Hot Cores}

\begin{figure*}
 \begin{center}
  \includegraphics[width=16cm]{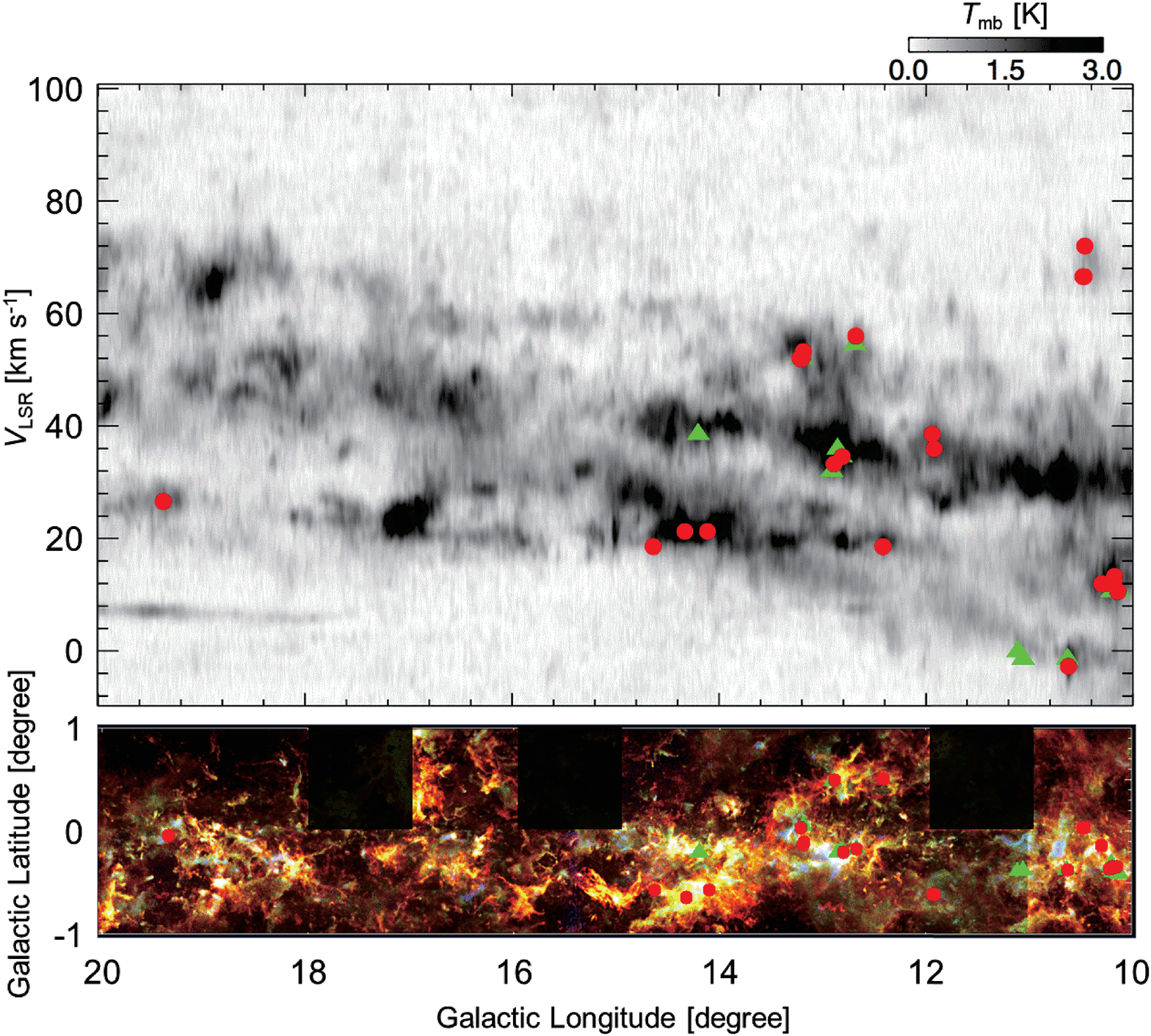} 
 \end{center}
\caption{Hot Cores and Dense Clumps distribution on FUGIN $^{13}$CO longitude-velocity diagram and CO image. \\
Upper panel: Hot Cores (red circles) and Dense Clumps (green triangles) distribution on $^{13}$CO longitude-velocity diagram.\\
Lower panel: Hot Cores and Dense Clumps distribution on the FUGIN three-color image of CO peak intensities (red:$^{12}$CO; green:$^{13}$CO; blue: C$^{18}$O) (color online). Three shaded squares indicate the regions where we did not extract hot core candidates due to bad signal-to-noise ratio. Noise level is different in every square as shown in figure \ref{fig:rms_map}.}
\label{fig:hc_region}
\end{figure*}

Figure \ref{fig:hc_region} shows distribution of Hot Cores and Dense Clumps overlaid on the FUGIN three-color image of the $^{12}$CO/$^{13}$CO/C$^{18}$O peak intensities and the $^{13}$CO $l$--$V$ diagram. 
We immediately note that many of the identified Hot Cores and Dense Clumps are clustered in space and velocity. 
Table \ref{tab:group} shows the 10 groups with more than two members with their distances (see section 3.3) and spatial extent. 
The groups have 2--9 members with spatial extents of $\sim$ 1--10 pc and radial velocity extent of $\sim$ 1--5 km s$^{-1}$. 
The spatial extent and velocity spreads are consistent with the size-linewidth relationship found in the interstellar medium (\cite{lar81}). 
This supports the idea that the members belongs to the same molecular cloud and are not artifact due to the line-of-site projection. 
The groups listed in table \ref{tab:group} include 29 sources (81\%) out of the 36 sources including both Hot Cores and Dense Clumps. 
If we count only those identified as Hot Cores, 16 (73\%) out of 22 Hot Cores are in 6 groups. 
This indicates that the formation process of high-mass stars proceeds concurrently at multiple locations within this spatial scale of $\sim$ 1--10 pc, i.e., within a molecular cloud. 

\subsection{Distance estimates}

We estimate the distances in tables \ref{tab:hc_cat} and \ref{tab:group} in the following way. 
Eight of the Hot Cores are associated with maser sources with trigonometric parallax measurements: 
\citet{san14} for G10.473$+$0.028 and G10.623$-$0.380; \citet{sat14} for G11.920$-$0.616; \citet{imm13} for G12.680$-$0.176 and G12.810$-$0.204; \citet{xu11} for G12.888$+$0.490; \citet{sat10} for G14.330$-$0.644; and \citet{wu14} for G14.636$-$0.569. 
We adopt these measurements as the most reliable distance estimates for these sources. 
For the sources that belong to the same groups with the above sources, we adopt the same distances because the spatial extent of the groups is small compared with the typical accuracy of the distance estimates. 
In this way, we assign trigonometric parallax-based distances to 13 Hot Cores and 3 Dense Clumps. 

For the remaining sources, we adopt the Bayesian distance proposed by \citet{rei14} that provides probability distribution of distance that can explain the Galactic longitude and radial velocity of the source. 
Since the Bayesian distance is based on the kinematic distance, the method generally proposes two distances corresponding near- and far-side of the tangential point in the region analyzed here. Far distances are estimated as far as 13--15 kpc. In resolving this near/far ambiguity, we choose near distance because our survey is limited by the sensitivity of the FUGIN data, which make detection of distant objects difficult.
The G10.18$-$0.35 group (W31-South) and G10.300$-$0.144 (W31-North) have spectrophotometric distance estimates of 3.55 kpc and 2.39 kpc, respectively, based on the near infrared photometry of the embedded stellar clusters (\cite{moi11}). 
Our adopted distances agree with these estimates, considering the accuracies of the distance estimates; $\sim$ 1 kpc for the spectrophotometric method and $\sim$ 0.2 kpc for the Bayesian method. 
The negative velocity of the G11.11$-$0.39 group makes it tricky to determine its distance.  
In the longitude-velocity diagram of figure \ref{fig:hc_region}, the G11.11$-$0.39 group appears to be associated with the same arm-like structure along with the G10.62$-$0.38 group, which has a trigonometric parallax distance. 
Consequently, we attribute to it the Bayesian distance close to the trigonometric parallax-based distance of the G10.62$-$0.38 group. 

\begin{table*}
\begin{center}
\caption{Groups of the FUGIN Hot Cores and Dense Clumps\\}
\scriptsize
  \begin{tabular}{c|ccc|cc}  
\hline
Group Name\footnotemark[$*$] & Member & $V{}_{\rm{LSR}}$ & Category\footnotemark[$\dag$] & Distance\footnotemark[$\ddag$] & Spatial extent \\
 & & [km s$^{-1}$] & & [kpc] & $l$ [pc] $\times b$ [pc] \\ \hline
 G10.18$-$0.35 & G10.131$-$0.408 & 10.6 & DC & 3.12 $\pm$0.21 & $9.3 \times 5.5$ \\ 
 & G10.145$-$0.339 & 10.6 & HC && \\
 & G10.178$-$0.350 & 13.3 & HC && \\ 
 & G10.187$-$0.344 & 12.0 & HC && \\ 
 & G10.195$-$0.397 & 10.6 & DC && \\ 
 & G10.203$-$0.341 & 12.0 & DC && \\
 & G10.203$-$0.308 & 10.6 & DC && \\
 & G10.209$-$0.336 & 12.0 & DC && \\
 & G10.220$-$0.369 & 12.0 & HC && \\
\hline 
G10.47$+$0.03 &G10.465$+$0.034  & 72.0 &  HC & 8.55 $^{+\,0.63}_{-\,0.55}$&$2.4 \times 2.4$ \\
 & G10.473$+$0.028 & 66.6 & HC & \\
 & G10.481$+$0.036 & 66.6 & HC & \\
\hline
G10.62$-$0.38 & G10.623$-$0.380 & $-2.7$\phantom{:} & HC  & 4.95 $^{+\,0.51}_{-\,0.43}$ & $0.7 \times 1.7$ \\
 & G10.631$-$0.372 & $-1.4$\phantom{:} & DC & \\ 
\hline 
G11.11$-$0.39 & G11.059$-$0.372 & $-1.4$\phantom{:} & DC  & 4.91 $\pm$0.25 & $4.3 \times 1.4$ \\ 
 & G11.109$-$0.388 & \phantom{0}0.0 & DC & \\
\hline 
G11.92$-$0.62 & G11.920$-$0.616 & 36.0 & HC & 3.37 $^{+\,0.39}_{-\,0.32}$ & $1.1 \times 0.0$ \\
 & G11.939$-$0.616 & 38.6 & HC & \\
\hline
G12.68$-$0.18 & G12.680$-$0.176 & 56.0 & HC & 2.40 $^{+\,0.17}_{-\,0.15}$ & $0.2 \times 0.2$ \\
 & G12.685$-$0.171 & 54.6 & DC & \\
\hline
G12.81$-$0.20 & G12.810$-$0.204  & 34.6 & HC & 2.92 $^{+\,0.35}_{-\,0.28}$ & $2.3 \times 1.0$ \\
 & G12.816$-$0.185 & 34.6 & DC & \\ 
 & G12.855$-$0.201 & 36.0 & DC & \\ 
\hline
G12.89$+$0.49 & G12.888$+$0.490  & 33.3 & HC & 2.50 $^{+\,0.28}_{-\,0.23}$ & $1.0 \times 0.9$ \\
 & G12.911$+$0.487 & 32.0 & DC & \\
\hline
G13.19$-$0.11 & G13.185$-$0.109  & 53.3 & HC & 4.60 $\pm$0.23 &$1.1 \times 2.0$ \\
 & G13.199$-$0.134 & 52.0 & HC & \\
\hline
G13.22$+$0.03 & G13.202$+$0.057  & 52.0 & DC & 4.57 $\pm$0.22&$8.0 \times 3.1$ \\
 & G13.216$+$0.029 & 52.0 & HC & \\
\hline
  \end{tabular}
\label{tab:group}
\begin{tablenotes}
\footnotemark[] There are seven isolated sources which are not described in this table. \\
\footnotemark[$*$] Groups are named after the most luminous member in each group, but with fewer digits for the coordinates. \\
\footnotemark[$\dag$] HC = Hot Core; DC = Dense Clump.\\
\footnotemark[$\ddag$] We assume all sources in a group are at the same distance. See section 3.3 for the methods of distance determination. \\
\end{tablenotes}
\end{center}
\end{table*}

\begin{figure}
 \begin{center}
  \includegraphics[width=8cm]{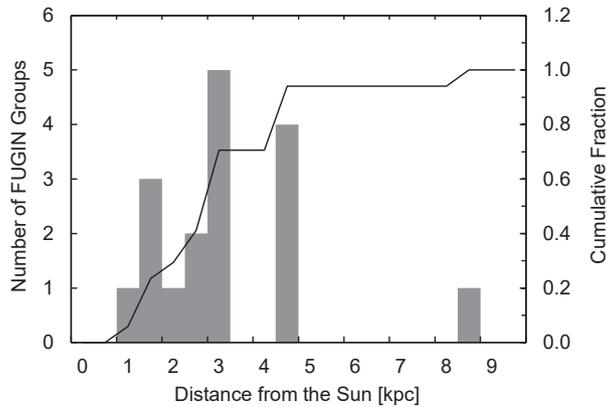} 
 \end{center}
\caption{Distance to the Hot Cores and Dense Clumps from the Sun. Its cumulative fraction is also shown. We count a group of sources as one entity along with isolated sources.}
\label{fig:dist_hist}
\end{figure}

Figure \ref{fig:dist_hist} shows the distribution of the distances. 
Here we count a group of sources as one entity along with isolated sources. 
The histogram shows that most of the sources identified in this study are within 5 kpc from the Sun, with a typical distance of 3 kpc. 
This may represent the sensitivity limit of our hot core survey. 
One notable exception is the G10.47$+$0.03 group at 8.55 $^{+ 0.63}_{-0.55}$ kpc (\cite{san14}), where we detect 3 very luminous Hot Cores. 

\section{Discussion}

\subsection{Nature of the FUGIN Hot Cores}

What is the nature of the FUGIN Hot Cores and Dense Clumps identified in this work?  
In this section we try to characterize their nature particularly in the context of high-mass star formation.

\subsubsection{Cross-matching with other signposts of high-mass star formation}

First, we compare our list of the FUGIN Hot Cores and Dense Clumps with the published catalogs of signposts of high-mass star formation. 
Table \ref{tab:crossmatch} shows the result of cross-matching of our catalog with the catalogs listed below. \\
{\bf{ATLASGAL}} (The APEX Telescope Large Area Survey of the Galaxy) clumps: Clumps have been identified from the 870 $\mu$m unbiased survey of the inner Galaxy with APEX and then followed up with millimeter wave spectroscopic observations with a variety of telescopes (\cite{urq18}). 
The list provides a large and systematic inventory of all massive, dense clumps in the Galaxy ($\ge 1000 M_\odot$ at a heliocentric distance of 20 kpc) and includes representative samples of all of the earliest embedded stages of high-mass star formation.
The original ATLASGAL survey was made with the APEX 12-m telescope with a $19.4''$ beam and a pointing accuracy of $\sim 4''$ rms (\cite{shu09}). \\
{\bf{Hi-GAL}} (Herschel infrared Galactic Plane Survey): Compact sources in 70 $\mu$m are taken from the catalog presented by \citet{mol16}. 
The Herschel satellite has a $6.0''$ diffraction-limited beam at 70 $\mu$m, and the astrometric accuracy of the compact source catalog is estimated as $\sim 1''$. \\
{\bf{MMB}} (Methanol Multibeam Survey): This is a 6.7 GHz methanol maser survey with a positional accuracy of $\sim$ $0.4''$ (\cite{gre10}) and a 12.2 GHz follow-up survey (\cite{bre14}). 
The observed maser lines are infrared pumped (Class II) and are expected in methanol-rich hot molecular gas illuminated by intense infrared radiation in high-mass star forming regions.\\
{\bf{WISE H\,\emissiontype{II}} regions}: This catalog of Galactic H\,\emissiontype{II} regions  has been made from the all-sky mid-infrared (12 $\mu$m and 22 $\mu$m) data from the WISE satellite and cross comparison with radio continuum surveys in the literature (\cite{and14}). 
The angular resolution of the WISE images is $6.5''$ and $12''$ for 12 $\mu$m and 22 $\mu$m, respectively. 
The catalog contains sources with wide range of sizes from $6''$ to $> 1000''$. \\
{\bf{CORNISH H\,\emissiontype{II}}} catalog: This is an unbiased 5 GHz radio continuum survey of the inner Galactic plane using the Very Large Array coordinated with the Spitzer GLIMPSE survey (\cite{hoa12}; \cite{pur13}). 
The survey detects emission on size scales from $1.5''$ up to $14''$, thus emphasizes more compact H\,\emissiontype{II} regions compared with the WISE catalog. 
We refer to the list of ``high reliability'' ($> 7\sigma $) sources identified as ultracompact H\,\emissiontype{II} regions or H\,\emissiontype{II} regions. \\
{\bf{CS survey}}: This is a CS $J=2-1$ survey for dense molecular gas associated with the IRAS point sources with color characteristics of ultracompact H\,\emissiontype{II} regions (\cite{bro95}). 
The observations were made with the OSO 20-m and SEST 15-m telescopes with angular resolutions of $39''$ and $50''$, respectively. \\
{\bf{EGO}} (A Catalog of Extended Green Objects): A survey at 3.6 $\mu$m, 4.5 $\mu$m, 5.8 $\mu$m, 8.0 $\mu$m, and 24 $\mu$m for extended 4.5 $\mu$m objects (\cite{cyg08}). Extended 4.5 $\mu$m emission may trace outflow and EGO sources correspond to high-mass star forming regions. The spatial resolution is within $2''$ in all bands.\\

\begin{figure}
 \begin{center}
  \includegraphics[width=8cm]{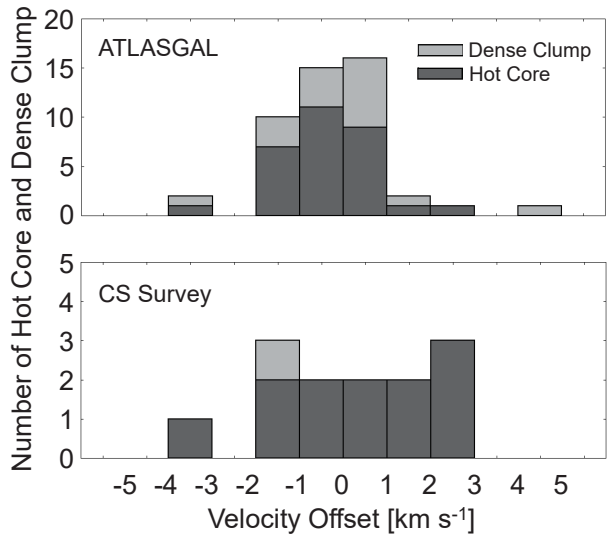} 
 \end{center}
\caption{The histogram of velocity offset between our sources and other catalog sources. Upper panel shows result with ATLASGAL (\cite{urq18}) and lower panel shows result with CS sources (\cite{bro95}).}
\label{fig:sokudosa}
\end{figure}

When the objects identified by various observational probes have a range of scale sizes, their cross matching is not straightforward in general. 
This is true in our situation. 
We identify the counterparts with the following criteria: 
(i) Positional agreement at the level of $< 30''$ (0.44 pc at 3 kpc) for Hi-GAL, CH$_3$OH maser and CORNISH sources, and $< 60''$ for the other sources. 
(ii) Source compactness at the level of $10'$. This is because we are interested in closer spatial coincidence and physical relationship with our sources (see section 3.1). 
(iii) Radial velocities coincident within $\approx 5$ km s$^{-1}$ when available. 
Figure \ref{fig:sokudosa} shows the offsets of the peak velocities of the identified ATLASGAL and CS survey sources. 
Median values of velocity offset are $-0.2$ km s$^{-1}$ for ATLASGAL and $-0.3$ km s$^{-1}$ for CS sources. 
In the case of MMB methanol masers, we require the peak velocity of our sources to fall within the velocity range of the maser emission.

ATLASGAL clumps are often extended and overlap with each other, and this makes it difficult to uniquely identify the counterparts of FUGIN Hot Cores from the coordinates only.  
In such cases we refer to the ATLASGAL image and select clumps that show better match with the Hot Core by eye inspection. 
We find ATLASGAL clumps  ON or NEAR 21 (95\%) of the 22 FUGIN Hot Cores. 
In the case of Hot Cores G10.465$+$0.034 and G10.473$+$0.028, we select the clump AGAL10.462$+$0.031 for the counterpart of both Hot Cores because the two Hot Cores appear to be embedded in the single continuous ATLASGAL clump.

Hi-GAL 70 $\mu$m sources are found ON or NEAR 17 (77\%) of the FUGIN Hot Cores. 

As for the association with H\,\emissiontype{II} regions, 16 (73\%) of the FUGIN Hot Cores have their H\,\emissiontype{II} region counterparts as indicated by WISE, CORNISH, and/or IRAS/UCH\,\emissiontype{II} CS catalogs. 
The CORNISH catalog provides a list of ultracompact and compact H\,\emissiontype{II} regions that are in more direct physical interaction with the Hot Cores, while the WISE catalog includes more extended H\,\emissiontype{II} regions showing the physical environment surrounding the Hot Cores as well. 
We find 6 Hot Cores (27\%) directly associated with the CORNISH H\,\emissiontype{II} regions. 
Methanol masers at 6.7 GHz are found ON or NEAR 7 Hot Cores (32\%). 
These coincidences along with the association with the Hi-GAL 70 $\mu$m sources suggest that these FUGIN Hot Cores are similar in nature to the classical hot cores associated with massive star formation. 

At the same time, we note that many Hot Cores do not show clear association with H\,\emissiontype{II} regions. 
Some Hot Cores may be associated with EGOs found nearby (G14.330$-$0.644, G14.636$-$0.569, G19.367$-$0.041), which may hint earlier stages of star formation or formation of less massive stars resulting in radio-quiet infrared sources. 
We also note that some Hot Cores are in the mini-starburst region W31 (G10.145$-$0.339, G10.178$-$0.350, G10.187$-$0.344, G10.220$-$0.369), and confusion with structures in the H\,\emissiontype{II} region complex could have hindered identifications of associated compact H\,\emissiontype{II} regions. 

For the remaining 13 candidates categorized as  FUGIN Dense Clumps, we find 11 ATLASGAL clumps ON or NEAR the position of the Dense Clumps (85\%), but they exhibit generally much less coincidence with the other signposts of massive star formation compared with the FUGIN Hot Cores. 

\begin{table*}
\begin{center}
\tbl{Correspondence of Hot Cores and Dense Clumps with other signposts of high-mass star formation.\footnotemark[$*$]\\ }{%
\scriptsize
  \begin{tabular}{c|ccccccc|c}  
\hline
FUGIN & ATLASGAL & Hi-GAL & Methanol & WISE H\,\emissiontype{II} & CORNISH & IRAS/UCH\emissiontype{II} & EGO\footnotemark[$\dag\dag$] &$L_{\rm bol}/M_{\rm clump}$ \\
 & clump\footnotemark[$\dag$] & $70 \mu$m\footnotemark[$\ddag$] & maser\footnotemark[$\S$] & region\footnotemark[$\|$] & 5 GHz\footnotemark[$\sharp$] & CS $J=2-1$\footnotemark[$**$] & & \\
\hline
{\bf Hot Cores} & & & & & & &\\
G10.145$-$0.339 & ON & ON & --- & --- & --- & --- & --- &195\phantom{.00} \\
G10.178$-$0.350 & ON & ON & --- & --- & --- & --- & --- &126\phantom{.00} \\
G10.187$-$0.344 & ON & NEAR & --- & --- & --- & --- & --- & 34.5\\
G10.220$-$0.369 & ON & --- & --- & --- & --- & --- & --- & 21.8\\
G10.300$-$0.144 & ON & ON & ON & --- & ON & NEAR & --- &68.1\\
G10.465$+$0.034 & ON & ON & --- & ON & ON & ON & --- & \phantom{00}46.7\footnotemark[$\ddag\ddag$] \\
G10.473$+$0.028 & NEAR & ON & ON & ON & ON & ON & --- & \phantom{00}46.7\footnotemark[$\ddag\ddag$] \\
G10.481$+$0.036 & --- & ON & ON  & ON & --- & NEAR & --- & --- \\
G10.623$-$0.380 & ON & ON & NEAR & ON & ON & --- & --- & 60.2\\
G11.920$-$0.616 & ON & ON & --- & NEAR & --- & NEAR & ON & \phantom{00}8.18\\
G11.939$-$0.616 & ON & ON & ON & ON & ON & ON & --- & 32.8\\
G12.416$+$0.510 & ON & NEAR & --- & ON & --- & ON & NEAR & 18.5\\
G12.680$-$0.176 & ON & NEAR & NEAR & NEAR & --- & --- & ON & \phantom{00}9.20\\
G12.810$-$0.204 & ON & ON & --- & ON & ON & --- & --- & --- \\
G12.888$+$0.490 & ON & ON & ON & ON & --- & ON & --- & 14.3\\
G13.185$-$0.109 & ON & ON & --- & ON & --- & --- & --- & \phantom{00}1.78\\
G13.199$-$0.134 & NEAR & --- & --- & --- & --- & ON & --- & 20.8\\
G13.216$+$0.029 & NEAR & --- & --- & --- & --- & --- & --- & --- \\
G14.113$-$0.569 & ON & ON & --- & NEAR & --- & --- & ---& \phantom{000}0.931 \\
G14.330$-$0.644 & ON & ON & --- & NEAR & --- & NEAR & NEAR & \phantom{00}8.07\\
G14.636$-$0.569 & NEAR & --- & --- & --- & --- & --- & NEAR & \phantom{00}3.54\\
G19.367$-$0.041 & NEAR & --- & --- & NEAR & --- & --- & NEAR & \phantom{00}4.09\\ 
\hline
{\bf Dense Clumps} & & & & & & &\\
G10.131$-$0.408 & ON & --- & --- & --- & --- & --- & --- & \phantom{00}3.62\\
G10.195$-$0.397 & NEAR & NEAR & --- & NEAR & --- & --- & --- & 20.3\\
G10.203$-$0.341 & ON & --- & ON & NEAR & --- & --- & --- & 21.8\\
G10.203$-$0.308 & NEAR & --- & --- & --- & --- & --- & --- & 14.4\\
G10.209$-$0.336 & NEAR & --- & --- & NEAR & --- & --- & --- & 21.7\\
G10.631$-$0.372 & NEAR & ON & --- & ON & --- & --- & --- & 60.3\\
G11.059$-$0.372 & --- & --- & --- & --- & --- & --- & --- & --- \\
G11.109$-$0.388 & NEAR & --- & --- & NEAR & NEAR & NEAR & --- & \phantom{00}9.71\\
G12.685$-$0.171 & NEAR & --- & --- & NEAR & --- & --- & NEAR & \phantom{00}9.20\\
G12.816$-$0.185 & --- & --- & --- & --- & --- & --- & --- & --- \\
G12.855$-$0.201 & ON & --- & --- & --- & --- & --- & --- & \phantom{00}2.52\\
G12.911$+$0.487 & ON & NEAR & --- & ON & --- & --- & --- & \phantom{00}7.78\\
G13.202$+$0.057 & --- & NEAR & --- & --- & --- & --- & --- & --- \\
G14.202$-$0.199 & ON & --- & --- & --- & --- & --- & --- & \phantom{00}1.19\\
\hline
\end{tabular}}
\label{tab:crossmatch}
\begin{tabnote}
\footnotemark[$*$] ON = correspondence within $20''$ for the Hi-GAL sources and methanol masers, and within $30''$ for the other sources; NEAR = correspondence with a position error of $20''$--$30''$ for the Hi-GAL sources and methanol masers, and $30''$--$60''$ for the other sources. \\
\footnotemark[$\dag$] ATLASGAL 870 $\mu$m clumps with molecular line follow-ups (\cite{urq18}).\\
\footnotemark[$\ddag$] 70 $\mu$m compact sources with $> 50$ Jy in the Herschel Hi-GAL survey (\cite{mol16}). \\
\footnotemark[$\S$] Class II methanol masers found in the 6.7 GHz Methanol Multibeam Survey (\cite{gre10}).\\
\footnotemark[$\|$] Galactic H\,\emissiontype{II} regions identified in mid-infrared by WISE (\cite{and14}).\\
\footnotemark[$\sharp$] ``High reliability'' 5 GHz sources identified as ultracompact H\,\emissiontype{II} regions or H\,\emissiontype{II} regions (\cite{pur13}).\\
\footnotemark[$**$] CS $J=2-1$ sources detected toward the IRAS point sources with color characteristics of ultracompact H\,\emissiontype{II} regions (\cite{bro95}).\\
\footnotemark[$\dag\dag$] Extended 4.5 $\mu$m object (\cite{cyg08}). \\
\footnotemark[$\ddag\ddag$] Same ATLASGAL clump (AGAL010.462$+$0.031) are selected as counterpart of two FUGIN Hot Cores.
\end{tabnote}
\end{center}
\end{table*}

\subsubsection{Evolutionary stage of the FUGIN Hot Cores}

We saw that many of the FUGIN Hot Cores appears to show close relationships with high-mass star formation. 
Now we discuss their evolutionary stages.

From the studies of the ATLAGAL-selected massive clumps, \citet{gia17} has shown that the ratio of bolometric luminosity to the clump mass statistically correlates with the evolutionary stage of the clump. 
The bolometric luminosity over clump mass ($L_{\rm bol}/M_{\rm clump}$) increases as the star formation stage evolves. 
The sources with $L_{\rm bol}/M_{\rm clump} \ < \ 2 \ L_\odot/M_\odot$ are considered as still accumulating sources,  those with $2 \ L_\odot/M_\odot \leq L_{\rm bol}/M_{\rm clump} \leq 40 \ L_\odot/M_\odot$ as Young Stellar Objects (YSOs) or Zero Age Main Sequences (ZAMS) stars and hot cores, and those with $L_{\rm bol}/M_{\rm clump} \ > \ 40 \ L_\odot/M_\odot$ as H\,\emissiontype{II} regions.

\begin{figure}
 \begin{center}
  \includegraphics[width=8cm]{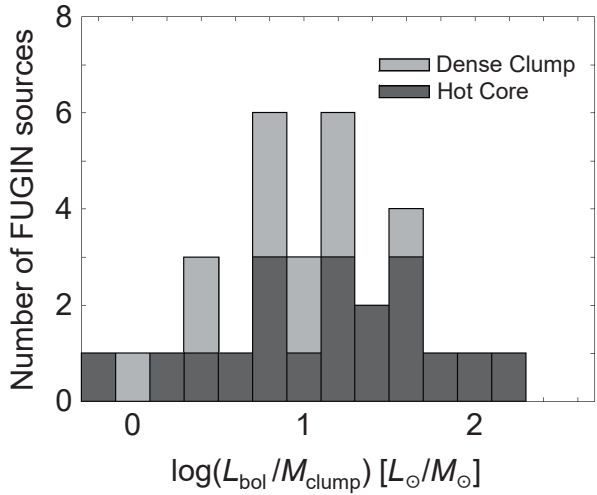} 
 \end{center}
\caption{The histogram of $L_{\rm bol}/M_{\rm clump}$ for FUGIN Hot Cores and Dense Clumps.}
\label{fig:lmhist}
\end{figure}

Figure \ref{fig:lmhist} shows the distribution of the $L_{\rm bol}/M_{\rm clump}$ ratios of the FUGIN Hot Cores and Dense Clumps. 
The $L_{\rm bol}/M_{\rm clump}$ are taken from the values for the corresponding ATLASGAL clumps listed in \citet{urq18}.
Eleven out of the 21 FUGIN Hot Cores (52\%) exhibit the $L_{\rm bol}/M_{\rm clump}$ ratio between 2 and 40 $L_\odot/M_\odot$, suggesting that they are statistically in the stage of YSOs or ZAMS stars and hot cores. 
This is in agreement with our conclusion that about 1/3 to 2/3 of the FUGIN Hot Cores are similar in nature to the classical hot cores. 
On the other hand, the larger spread in the $L_{\rm bol}/M_{\rm clump}$ distribution beyond the $2 - 40 L_\odot/M_\odot$ range suggests that the FUGIN Hot Cores may also include objects in evolutionary stages earlier or later than that of the classical hot cores. 

\subsection{Limitations of this survey}

As described in section 2.1, our survey of hot cores starts from finding source candidates in the FUGIN data. 
Since the sensitivity of the FUGIN survey has been set for its original purpose of the Galactic survey in the $^{12}$CO, $^{13}$CO and C$^{18}$O lines, the signal to noise ratios for the weaker HNCO and CH$_3$CN lines we search as the hot core signatures are not very high.  
The noise varies typically by a factor of two from a square degree region to another due to the varying observing conditions and available time during the original FUGIN observations. 
In addition, the stability of the four receivers corresponding to the four beams of FOREST was uneven during the observations, leaving strong scanning patterns in the maps in some regions. 
Here we examine the limitations and cautions in the use of the present results caused by this non-ideal situation. 

Since we have set the detection limit for the stacked HNCO and CH$_3$CN lines at $5\sigma$ (0.1--0.2 K) referring to the noise level of the individual stacked spectra\footnote{By doing so, we can reduce false candidate detection caused by the scanning effect, because the spectra with the scanning effect have higher rms noise.}, 
the sensitivity limit for the candidate sources is uneven. 
Nevertheless, we report all the detected Hot Cores and Dense Clumps in this paper because it would have more merit of increasing the number of sources at this stage of the research. 
We can impose additional selection criteria to get a subset from the present list if our analysis requires a sample with better-controlled sensitivity limit. 

The positional accuracy of the sources is also limited by the relatively low signal to noise ratio. 
The maps we use in the candidate selection have been made with $10''$ grid with a gridding convolution function that corresponds to a HPBW of $\approx 20''$ (\cite{saw08}). 
This limit the positional accuracy for the first place. 
When the signal to noise ratio is low, the noise fluctuation may cause positional error of one pixel ($10''$) in $l$ or $b$. 
When we have two or more detections at spatially adjacent pixels, we take the position of the strongest intensity of the stacked HNCO and CH$_3$CN lines as the position of the candidate. 
This can introduce further positional error if the source is actually a marginally resolved cluster (i.e., source confusion) or the intensity distribution is distorted by the noise. 
By these limitations, we expect that the errors of the source position are about $10''$ with some cases getting even larger. 

Among the objects used in the cross-matching in section 4.1.1, the Hi-GAL 70 $\mu$m source and the MMB 6.7 GHz methanol masers are compact in size and have high positional accuracy of the order of an arcsecond. 
Out of the 17 Hi-GAL 70 $\mu$m sources exceeding 50 Jy found within $30''$ from the nominal position of the FUGIN Hot Cores, 7 (41\%) and 14 (82\%, cumulative) are within $10''$ and $20''$ of the Hot Core positions, respectively. 
Similarly, out of the 7 MMB methanol maser sources, 3 (43\%) and 5 (71\%, cumulative) are found within $10''$ and $20''$ of the Hot Core positions, respectively. 
These counts may overestimate the scatter of the positional error, because the measured values reflect projected distance, and because the counts may be slightly contaminated by foreground and background sources in the Galactic disk by chance coincidence. 
This is consistent with our expectations that the positional accuracy of the FUGIN Hot Cores is about $10''$ with possible errors larger than $20''$ for some sources. 

As a result of the uneven sensitivity limit and positional errors that may also have affected the confirmation observations, we have missed some real sources. 
For example,  in the ATLASGAL-based spectral-line follow-up (Millimetre Astronomy Legacy Team 90 GHz (MALT90) survey), \citet{rat16} detected 7 sources of CH$_3$CN emission exceeding $T_{\rm A}^* = 0.15$ K with the Mopra 24-m telescope in the area of the present survey. 
Four of them are identified as Hot Cores in our survey, while we missed 3 sources; AGAL12.904$-$0.031, AGAL13.178$+$0.059 and AGAL14.777$-$0.487. 
In using the FUGIN Hot Core survey results, we need to keep in mind the possibility of missed sources.

\section{Summary}

We have developed a method to survey hot cores without bias to continuum emission by using the FUGIN (FOREST Unbiased Galactic plane Imaging survey with the Nobeyama 45-m telescope) data. 
We first listed hot core candidates based on the C$^{18}$O $J=1-0$ and the stacked HNCO $J_{K_a, K_c}=5_{0,5}-4_{0,4}$ plus CH$_3$CN $J=6-5$ line intensities and their distribution in the FUGIN datacube. 
We then conducted more sensitive follow-up observations of the candidates in hot core tracer lines by using the Nobeyama 45-m telescope to confirm and characterize the identified objects. 
The observed lines include C$^{34}$S $J=2-1$, SO $J_K=2_3-1_2$, OCS $J=9-8$, HC$_3$N $J=12-11$, CH$_3$OH $J=2-1$ and $J=8-7$, in addition to the C$^{18}$O $J=1-0$, HNCO $J_{K_a, K_c}=5_{0,5}-4_{0,4}$ and CH$_3$CN $J=6-5$ lines used in the candidate selection. 
We define FUGIN Hot Cores as sources of spatially compact ($\ll 100''$) emission detected above $3\sigma$ in at least two of the hot core tracer lines. 
We also detect Dense Clumps which do not meet the Hot Core criteria but have at least two lines that indicate high molecular gas density. 

We have applied this method to the 17 square degrees area in the $l =10^\circ$--$20^\circ$ region of the FUGIN survey, and identified 22 Hot Cores and 14 Dense Clumps. 
The main findings from the initial analysis of these sources are summarized as follow: 
\begin{enumerate}
  \item Most of the identified Hot Cores and Dense Clumps are within 5 kpc from the Sun with a typical distance of 3 kpc. 
  This may represent the sensitivity limit of the survey. 
  \item Many of the sources are clustered in groups of $\sim$ 1--10 pc in diameter, showing that formation of high-mass stars proceeds concurrently at multiple locations within this spatial scale. 
  The spatial extent and the velocity spread of cores within each group is roughly in agreement with the size-linewidth relation of interstellar molecular clouds (\cite{lar81}).  
  \item Many of the FUGIN Hot Cores have counterparts in ATLASGAL dust clumps (95\%), WISE H\,\emissiontype{II} regions (64\%), and Class II methanol masers (32\%). 
  This suggests that the FUGIN Hot Core sample is closely related to regions of high-mass star formation. 
  \item For the Hot Cores with ATLASGAL counterparts, we find their bolometric luminosity to clump mass ratios, $L_{\rm bol}/M_{\rm clump}$, which range from $\sim 1 \ L_\odot/M_\odot$ to $\sim 10^2 \ L_\odot/M_\odot$. 
  About $2/3$ of them fall within the range of 2--40 $L_\odot/M_\odot$ that statistically corresponds to the stages of the accreting protostars, ZAMS stars and hot cores (\cite{gia17}). 
The FUGIN Hot Core sample may cover somewhat broader range of star formation stages than the classical hot core phase. 
\end{enumerate}

On the basis of the results described in this paper, we are expanding the analysis and corresponding confirmation observations to the $l = 20^\circ$--$50^\circ$ part of the FUGIN data, with an aim of completing the catalog of FUGIN Hot Cores for the first Galactic quadrant. 
This would significantly increase the number of Hot Cores available for the statistical analysis of their evolution and chemistry.  
The catalog would also provide a useful starting point for high resolution studies of physical and chemical structures of a sample of high-mass star forming regions with ALMA. 

\begin{ack}

We thank Drs. Hideko Nomura, Yuri Aikawa and Satoshi Yamamoto for helpful discussions concerning the chemical abundance of hot cores. 
This publication makes use of data from FUGIN, FOREST Unbiased Galactic plane Imaging survey with the Nobeyama 45-m telescope, a legacy project in the Nobeyama 45-m radio telescope. 
The Nobeyama 45-m radio telescope is operated by Nobeyama Radio Observatory, a branch of National Astronomical Observatory of Japan. 
Data analysis was in part carried out on the Multi-wavelength Data Analysis System operated by the Astronomy Data Center (ADC), National Astronomical Observatory of Japan. 

\end{ack}


\begin{thebibliography}{99}

\bibitem[Anderson et al.(2014)]{and14} Anderson, L.~D., Bania, T.~M., Balser, D.~S., Cunningham, V., Wenger, T.~V., Johnstone, B.~M.; Armentrout, W.~P.\ 2014, \apjs, 212, 1

\bibitem[Bisschop et al.(2007)]{bis07} Bisschop, S.~E., J{\o}rgensen, J.~K., van Dishoeck, E.~F., \& de Wachter, E.~B.~M.\ 2007, \aap, 465, 913

\bibitem[Breen et al.(2014)]{bre14} Breen, S.~L., Ellingsen, S.~P., Caswell, J.~L., et al.\ 2014, \mnras, 438, 3368 

\bibitem[Bronfman et al.(1996)]{bro95} Bronfman, L., Nyman, L.-\AA., \& May, J.\ 1996, \aaps, 115, 81 

\bibitem[Chen et al.(2011)]{che11} Chen, X., Ellingsen, S.~P., Shen, Z.-Q., Titmarsh, A., \& Gan, C.-G.\ 2011, \apjs, 196, 9 

\bibitem[Cyganowski et al.(2008)]{cyg08} Cyganowski, C.~J., Whitney, B.~A., Holden, E., et al.\ 2008, \aj, 136, 2391 


\bibitem[Giannetti et al.(2017)]{gia17} Giannetti, A., Leurini, S., Wyrowski, F., Urquhart, J., Csengeri, T., Menten, K.~M.; K\"onig, C., \& G\"usten, R.\ 2017, \aap, 603, A33

\bibitem[Green et al.(2010)]{gre10} Green, J.~A., Caswell, J.~L., Fuller, G.~A., et al.\ 2010, \mnras, 409, 913 

\bibitem[Hoare et al.(2012)]{hoa12} Hoare, M.~G., Purcell, C.~R., Churchwell, E.~B., et al.\ 2012, \pasp, 124, 939

\bibitem[Immer et al.(2013)]{imm13} Immer, K., Reid, M.~J., Menten, K.~M., Brunthaler, A., \& Dame, T.~M.\ 2013, \aap, 553, A117 

\bibitem[Kuno et al.(2011)]{kun11} Kuno, N., Takano, S., Iono, D., et al.\ 2011, XXXth URSI General Assembly and Scientific Symposium, JP2-19

\bibitem[Larson(1981)]{lar81} Larson, R.~B.\ 1981, \mnras, 194, 809 

\bibitem[Minamidani et al.(2016)]{min16} Minamidani, T., Nishimura, A., Miyamoto, Y., et al.\ 2016, \procspie, 9914, 99141Z

\bibitem[Mois{\'e}s et al.(2011)]{moi11} Mois{\'e}s, A.~P., Damineli, A., Figuer{\^e}do, E., et al.\ 2011, \mnras, 411, 705 

\bibitem[Molinari et al.(2016)]{mol16} Molinari, S., Schisano, E., Elia, D., et al.\ 2016, \aap, 591, A149 


\bibitem[Nomura \& Millar(2004)]{nom04} Nomura, H., \& Millar, T.~J.\ 2004, \aap, 414, 409 


\bibitem[Purcell et al.(2013)]{pur13} Purcell, C.~R., Hoare, M.~G., Cotton, W.~D., et al.\ 2013, \apjs, 205, 1

\bibitem[Rathborne et al.(2016)]{rat16} Rathborne, J.~M., Whitaker, J.~S., Jackson, J.~M., et al.\ 2016, Publ.\ Astron.\ Soc.\ Aust, 33, e030

\bibitem[Reid et al.(2014)]{rei14} Reid, M.~J., Menten, K.~M., Brunthaler, A., et al.\ 2014, \apj, 783, 130

\bibitem[Sanna et al.(2014)]{san14} Sanna, A., Reid, M.~J., Menten, K.~M., et al.\ 2014, \apj, 781, 108 

\bibitem[Sato et al.(2010)]{sat10} Sato, M., Hirota, T., Reid, M.~J., et al.\ 2010, \pasj, 62, 287 

\bibitem[Sato et al.(2014)]{sat14} Sato, M., Wu, Y.~W., Immer, K., et al.\ 2014, \apj, 793, 72 


\bibitem[Sawada et al.(2008)]{saw08} Sawada, T., Ikeda, N., Sunada, K., et al.\ 2008, \pasj, 60, 445 

\bibitem[Schuller et al.(2009)]{shu09} Schuller, F., Menten, K.~M., Contreras, Y., et al.\ 2009, \aap, 504, 415 

\bibitem[Umemoto et al.(2017)]{ume17} Umemoto, T., Minamidani, T., Kuno, N., et al.\ 2017, \pasj, 69, 78 

\bibitem[Urquhart et al.(2018)]{urq18} Urquhart, J.~S., K{\"o}nig, C., Giannetti, A., et al.\ 2018, \mnras, 473, 1059 


\bibitem[Wu et al.(2014)]{wu14} Wu, Y.~W., Sato, M., Reid, M.~J., et al.\ 2014, \aap, 566, A17 

\bibitem[Xu et al.(2011)]{xu11} Xu, Y., Moscadelli, L., Reid, M.~J., et al.\ 2011, \apj, 733, 25 

\bibitem[Yamamoto(2017)]{yam17} Yamamoto, S. 2017, Introduction to Astrochemistry: Chemical Evolution from Interstellar Clouds to Star and Planet Formation (Tokyo: Springer)



\end{thebibliography}
\end{document}